\documentclass[onecolumn,prd,nofootinbib,aps,floats,floatfix,amsmath,amssymb,secnumarabic,preprintnumbers,showpacs,notitlepage]{revtex4-1} %
\usepackage[final]{graphicx}
\usepackage{amsmath}
\usepackage{amsfonts}
\usepackage{amssymb}
\usepackage{latexsym}
\usepackage{graphicx}
\usepackage[english]{babel}
\usepackage{multirow}
\usepackage{float}
\usepackage{url}
\usepackage{hyperref}
\usepackage{slashed}
\usepackage{xcolor}

\def\beq{\begin{equation}}
\def\eeq{\end{equation}}
\def\bea{\begin{eqnarray}}
\def\eea{\end{eqnarray}}
\def\nn{\nonumber}
\def\roughly#1{\mathrel{\raise.3ex\hbox
{$#1$\kern-.75em\lower1ex\hbox{$\sim$}}}}

\def\sla#1{\raise.15ex\hbox{$/$}\kern-.57em #1}
\def\BDtaunu{\bar{B} \to D^+ \tau^{-} {\bar\nu}_\tau}
\def\BDlnu{\bar{B} \to D^+ \ell^{-} {\bar\nu}_\ell}
\def\BDstartaunu{\bar{B} \to D^{*+} \tau^{-} {\bar\nu}_\tau}
\def\BDstarlnu{\bar{B} \to D^{*+} \ell^{-} {\bar\nu}_\ell}
\def \bbtosb{{\bar b}\to{\bar s}}

\def \({\left(}
\def \){\right)}

\def \l.{\left.}
\def \r.{\right.}
\def \nn{\nonumber}
\def \nl{\nn\\}
\def \s{\sqrt{2}}

\def \SM{{\rm SM}}

\def \expt{{\rm exp}}

\def \bwt{\begin{widetext}}
\def \ewt{\end{widetext}}
\def \bma{\begin{matrix}}
\def \ema{\end{matrix}}
\def \Wtaunu{$W$-$\tau$-$\nu_\tau$}

\begin{document}
\preprint{UdeM-GPP-TH-16-249, UMISS-HEP-2016-01}
\title{\boldmath Is there really a $W \to \tau\nu$ puzzle?}
\author{Bhubanjyoti Bhattacharya\footnote{bhujyo@lps.umontreal.ca}}
\affiliation{Physique des Particules, Universit\'e de Montr\'eal,
C.P. 6128, succ.\ centre-ville, Montr\'eal, Quebec, H3C 3J7, Canada}
\author{Alakabha Datta\footnote{datta@phy.olemiss.edu}}
\affiliation{
Department of Physics and Astronomy,
University of Mississippi,
Lewis Hall, University, Mississippi, 38677, USA}
\author{David London\footnote{london@lps.umontreal.ca}}
\affiliation{Physique des Particules, Universit\'e de Montr\'eal,
C.P. 6128, succursale centre-ville, Montr\'eal, Quebec, H3C 3J7, Canada}

\begin{abstract}
According to the Particle Data Group, the measurements of ${\cal B}
(W^+ \to \tau^+ \nu_\tau)$ and ${\cal B}(W^+ \to \ell^+ \nu_\ell)$
($\ell = e,\mu$) disagree with one another at the $2.3\sigma$ level.
In this paper, we search for a new-physics (NP) explanation of this $W
\to \tau\nu$ puzzle.  We consider two NP scenarios: (i) the $W$ mixes
with a $W'$ boson that couples preferentially to the third generation,
(ii) $\tau_{L,R}$ and $\nu_{\tau L}$ mix with isospin-triplet leptons.
Unfortunately, once other experimental constraints are taken into
account, neither scenario can explain the above experimental result.
Our conclusion is that the $W \to \tau\nu$ puzzle is almost certainly
just a statistical fluctuation.
\end{abstract}


\maketitle

\section{Introduction}

At present, there are a few measurements that are in potential
disagreement with the predictions of the standard model (SM) of
particle physics. One hint of lepton nonuniversality involves the
leptonic decays of the $W$. According to the Particle Data Group
\cite{pdg}, we have
\bea
{\cal B}(W^+ \to e^+ \nu_e) &=& (10.71 \pm 0.16) \% ~, \nn\\
{\cal B}(W^+ \to \mu^+ \nu_\mu) &=& (10.63 \pm 0.15) \% ~, \nn\\
{\cal B}(W^+ \to \tau^+ \nu_\tau) &=& (11.38 \pm 0.21) \% ~,
\label{direct}
\eea
yielding
\beq
\frac{2 {\cal B}(W^+ \to \tau^+ \nu_\tau)}{{\cal B}(W^+ \to e^+ \nu_e)
+ {\cal B}(W^+ \to \mu^+ \nu_\mu)} = 1.067 \pm 0.029 ~.
\eeq
The SM prediction for this ratio is 0.999 to a very good
approximation, so there is a difference at the level of $2.3\sigma$.
We refer to this as the ``$W \to \tau\nu$ puzzle.'' Of course, this
could simply be a statistical fluctuation. But could it in fact be due
to the presence of new physics (NP)?

In the past, the only theoretical studies that attempted to directly
address the $W \to \tau \nu$ puzzle involved models with two-Higgs
doublets. Specifically, it was suggested that the excess in
$W\to\tau\nu$ events is due to contamination by a light charged Higgs,
with mass $m_W$, decaying via $H^+ \to \tau^+ \nu_\tau$~
\cite{chargedHiggs}. However, recently the data of the four LEP
collaborations was combined and a search for pair-produced charged
Higgs bosons was performed \cite{LEPcomb}. No significant excess of
$\tau^+ \nu_\tau$ final states was observed compared to the SM
background, so that a lower limit can be set on the mass of the
charged Higgs as a function of the $H^+ \to\tau^+ \nu_\tau$ branching
ratio. While the LEP study does not completely rule out
two-Higgs-doublet models with the most general couplings, it does
severely restrict the available parameter space \cite{Cline:2015lqp}.

An alternative explanation of the puzzle is that the \Wtaunu\ coupling
is itself increased. This possibility was considered in
Ref.~\cite{EFT} using an effective field theory (EFT) approach. Here
the NP effects are encapsulated by including higher-dimensional
operators, each with its own arbitrary Wilson coefficient. The authors
study the effect of different flavor symmetries; they conclude that it
is difficult to resolve the \Wtaunu\ puzzle in this framework, mainly
due to the constraints arising from $Z$ and $\tau$ decays. However, we
note that this analysis is not the most general -- the question of
neutrino masses has not been considered. In the EFT, neutrino mass
operators arise at dimension 5. The authors of Ref.~\cite{EFT}
write, ``The only gauge-invariant operator of dimension five violates
lepton number, and thus it can be safely neglected under the
assumption that the violation of that symmetry occurs at scales much
higher than $\Lambda \sim 1$~TeV.'' But this assumption is not
necessarily true. Indeed, in the present paper we consider several
models giving rise to lepton-number violation at a scale of $O(1)$ TeV
(see Sec.~III), and generating nonzero neutrino masses. These are
correlated with the contribution to the \Wtaunu\ coupling. The
connection between neutrino masses and the \Wtaunu\ coupling is
ignored in the EFT approach, but is taken into account here.

A larger \Wtaunu\ coupling can also improve some other discrepancies
with the SM. Below we discuss several other measurements that are
sensitive to the \Wtaunu\ coupling. It should be noted that, while
Eq.~(\ref{direct}) is a direct measurement of the \Wtaunu\ coupling,
these other measurements are indirect probes of this coupling, and
there may be other new-physics contributions to these decays
\cite{datta}.

Consider first the decay $B^+ \to \tau^+ \nu_\tau$. Its branching
ratio is \cite{pdg}
\beq
{\cal B}(B^+ \to \tau^+ \nu_\tau) = (1.14 \pm 0.27) \times 10^{-4} ~,
\label{indirect1}
\eeq
while the SM prediction is \cite{AGC}
\bea
{\cal B}(B^+ \to \tau^+ \nu_\tau) & = & \tau_{B^+} G_F^2 m_\tau^2 f_B^2
|V_{ub}|^2 \frac{m_B}{8\pi} \left( 1 - \frac{m_\tau^2}{m_B^2}
\right)^2 \nn\\
&=& (0.81 \pm 0.08) \times 10^{-4} ~.
\label{BtaunuSMpred}
\eea
Here the FLAG average $f_B = (190.5 \pm 4.2)$~MeV~\cite{FLAG} and the
CKMfitter result $|V_{ub}| = (3.55 \pm 0.16) \times 10^{-3}$~
\cite{CKMfitter} have been used. From the above numbers, we see that
there is a small ($1.5\sigma$) disagreement between the measurement
and the SM prediction. It is stressed in Ref.~\cite{AGC} that the size
of the disagreement depends on the value taken for $|V_{ub}|$, and
there is a long-standing discrepancy between the determinations of
$|V_{ub}|$ from inclusive $B \to X_u \ell^+ \nu$ and exclusive ${\bar
  B} \to M \ell {\bar\nu}$ decays~\cite{pdg}. Indeed, if the inclusive
value for $|V_{ub}|$ is used, the disagreement disappears. Still, if
the SM prediction of Eq.~(\ref{BtaunuSMpred}) holds, the agreement
with experiments can be improved if the \Wtaunu\ coupling is increased.

Another example, similar to the above process, involves $D_s^+ \to
\tau^+ \nu_{\tau}$ and $D_s^+ \to \ell^+ \nu_{\ell}$ ($\ell = e, \mu$)
decays.  Experimentally, it is found that~\cite{pdg}
\bea
R_{D_s} \equiv \frac{{\cal B}( D_s^+ \to \tau^+ \nu_{\tau})}{{\cal B}
(D_s^+ \to \ell^+ \nu_{\ell})} = 10.0 \pm 0.6 ~.
\label{indirect3}
\eea
In the SM, this ratio is predicted to be
\beq
R_{D_s} = \frac{m^2_\tau}{m^2_\mu} \frac{(1 - m^2_\tau/m^2_{D^+_s})^2}
{(1 - m^2_\mu/m^2_{D^+_s})^2} = 9.742 \pm 0.013 ~.
\eeq
Due to the large experimental error on $R_{D_s}$, there is no
discrepancy with the SM, but, at the $3 \sigma$ level, a 10\% increase
in the \Wtaunu\ coupling is allowed.  This measurement is thus
consistent with that of Eq.~(\ref{direct}).

$\tau$ decays would obviously be affected by a change in the
\Wtaunu\ coupling. Consider first $\tau^- \to e^- \nu_\tau
                 {\bar\nu}_e$. Here the SM predicts \cite{pdg}
\beq
R_\tau \equiv \frac{{\cal B}(\tau^- \to e^- \nu_\tau {\bar\nu}_e)}
{{\cal B}(\mu^- \to e^- \nu_\mu {\bar\nu}_e)}
= \frac{\tau_\tau}{\tau_\mu}\(\frac{m_\tau}{m_\mu}\)^5
~=~ (17.77 \pm 0.03)\%~,
\eeq
where $\tau_i$ represents the mean lifetime of particle $i$. The
experimental value for the above ratio is
\beq
R_\tau \approx (17.83 \pm 0.04)\% ~,
\eeq
assuming the branching ratio for the $\mu$ decay is $\approx 100\%$
\cite{pdg}. This $\tau$ decay channel therefore allows very little
(less than a percent) change in the \Wtaunu\ coupling.

A second decay is $\tau^- \to \pi^- \nu_\tau$. In the SM, the
branching ratio for this decay can be expressed as \cite{barish}
\bea
{\cal B}(\tau^- \to \pi^- \nu_\tau) &=& \frac{G_F^2|V_{ud}|^2}
{16\pi}f_\pi^2\tau_\tau m_\tau^3 \left(1-\frac{m_\pi^2}{m_\tau^2}
\right)^2  \delta_{\tau/\pi} \nl
&=& (10.67\pm0.23)\% ~.
\label{taupinu}
\eea
Here $\delta_{\tau/\pi}$ represents the small radiative corrections to
the decay rate; it is known very well: $\delta_{\tau/\pi} = 1.0016 \pm
0.0014$ \cite{rad}. Above we have used the FLAG average $f_\pi =
(130.2 \pm 1.4)$~MeV~\cite{FLAG} and the CKMfitter result $|V_{ud}| =
(0.97425 \pm 0.00022)$ \cite{CKMfitter}. (The biggest source of the
$\sim 2\%$ error in the predicted branching ratio is the lattice value
for $f_\pi$ which has a $\sim 1\%$ error.) The prediction in
Eq.~(\ref{taupinu}) should be compared with the measured value
\cite{pdg}
\beq
{\cal B}(\tau^- \to \pi^- \nu_\tau) = (10.91 \pm 0.07)\% ~.
\eeq
In this case the predicted value has a larger error than the measured
value and they are consistent with each other. Still, if we allow for
a $3\sigma$ (upward) deviation from the measured value and trust the
predicted central value, we find that a 2\% increase in the
\Wtaunu\ coupling is allowed. Also, it must be remembered that this
decay can be affected by other new-physics contributions
\cite{datta_tau}.

Finally, there are the charged-current decays $\bar{B} \to D^{(*)+}
\ell^{-} {\bar\nu}_\ell$, which have been measured by the BABAR~
\cite{RD_BaBar}, Belle~\cite{RD_Belle}, and LHCb \cite{RD_LHCb}
collaborations. It is found that the values of the ratios ${\cal
  B}(\bar{B} \to D^{(*)+} \tau^{-} {\bar\nu}_\tau)/{\cal B}(\bar{B}
\to D^{(*)+} \ell^{-} {\bar\nu}_\ell)$ ($\ell = e,\mu$) considerably
exceed their SM predictions. The experimental results and theoretical
predictions can be combined to yield \cite{RDRK_Isidori}
\bea
R_D &\equiv& \frac{ {\cal B}(\BDtaunu)_{\expt} / {\cal B}(\BDtaunu)_{\SM} }
{ {\cal B}(\BDlnu)_{\expt} / {\cal B}(\BDlnu)_{\SM} } = 1.37 \pm 0.18 ~, \nn\\
R_{D^*} &\equiv& \frac{ {\cal B}(\BDstartaunu)_{\expt} / {\cal B}(\BDstartaunu)_{\SM} }
{ {\cal B}(\BDstarlnu)_{\expt} / {\cal B}(\BDstarlnu)_{\SM} } = 1.28 \pm 0.08 ~.
\label{indirect2}
\eea
The measured values of $R_D$ and $R_{D^*}$ represent deviations from
the SM of 2.0$\sigma$ and 3.8$\sigma$, respectively.  This is the
$R_{D^{(*)}}$ puzzle. In this case, the discrepancies with the SM are
too large to be explained entirely by an increase in the
\Wtaunu\ coupling. Even so, such an increase would lead to larger
theoretical predictions for ${\cal B}(\bar{B} \to D^{(*)+} \tau^{-}
{\bar\nu}_\tau)$, which would reduce the disagreement with experiments.

We therefore see that an increased \Wtaunu\ coupling can explain the
$W \to \tau\nu$ puzzle, and can also improve other discrepancies with
the SM. (It must also be conceded that not all measurements support
the idea of an increased coupling.) The purpose of this paper is to
attempt to find a NP model in which the \Wtaunu\ coupling can be made
larger.

To this end we consider two NP possibilities. In the first, we assume
that a $W'$ boson exists that couples preferentially to the third
generation. The mixing of this $W'$ with the SM $W$ could then lead to
an increased \Wtaunu\ coupling. In the second, we allow
the $\tau_{L,R}$ and $\nu_{\tau L}$ to mix with isospin-triplet leptons.
Once again, this mixing could generate a larger \Wtaunu\
coupling. Unfortunately, as we will see, once constraints from other
measurements are taken into account, neither NP scenario can reproduce
the measured \Wtaunu\ coupling. Because of the difficulty
in finding a reasonably simple NP explanation, we are forced to
conclude that the $W \to \tau\nu$ puzzle is probably just a
statistical fluctuation.

We begin in Sec.~II with an evaluation of the potential for $W$-$W'$
mixing to lead to an increased \Wtaunu\ coupling once all experimental
constraints are taken into account. This analysis is repeated in
Sec.~III for the mixing of the $\tau_{L,R}$ and $\nu_{\tau L}$ with
isospin-triplet leptons. (The details of the formalism of this mixing
are given in the Appendix.) We conclude in Sec.~IV.

\section{\boldmath $W$-$W'$ Mixing}

There has been another recent hint of lepton nonuniversality. The
LHCb Collaboration measured the ratio of decay rates for $B^+ \to K^+
\ell^+ \ell^-$ ($\ell = e,\mu$) in the dilepton invariant mass-squared
range 1~GeV$^2$ $\le q^2 \le 6$~GeV$^2$ \cite{RKexpt}, and found
\bea
R_K & \equiv & \frac{{\cal B}(B^+ \to K^+ \mu^+ \mu^-)}{{\cal B}(B^+ \to
  K^+ e^+ e^-)} \nn\\
&=& 0.745^{+0.090}_{-0.074}~{\rm (stat)} \pm 0.036~{\rm (syst)} ~.
\eea
This differs from the SM prediction of $R_K = 1 \pm O(10^{-4})$
\cite{RKtheory} by $2.6\sigma$. A NP explanation of this $R_K$ puzzle
was offered in Ref.~\cite{GGL}.  Here the NP is assumed to couple
preferentially to the third generation, giving rise to the
operator\footnote{The $(V-A)\times (V-A)$ form of this operator follows
  the analysis of Ref.~\cite{HS1}. There it is found that the only
  NP operator that can reproduce the experimental value of $R_K$ is
  $({\bar s}\gamma_\mu P_L b)({\bar \ell} \gamma^\mu P_L \ell)$. This
  is consistent with the NP explanations for the $B\to
  K^{(*)}\mu^+\mu^-$ angular distributions measured by LHCb
  \cite{HurthAS}.}
\beq
G ({\bar b}'_L \gamma_\mu b'_L) ({\bar \tau}'_L \gamma^\mu \tau'_L) ~,
\label{GGLoperator}
\eeq
where $G = O(1)/\Lambda_{NP}^2 \ll G_F$, and the primed fields are the
fermion eigenstates in the gauge basis. When one transforms to the
mass basis, this generates the operator $({\bar b}_L \gamma_\mu s_L)
({\bar \mu}_L \gamma^\mu \mu_L)$ that contributes to $\bbtosb \mu^+
\mu^-$. (There is also a contribution to $\bbtosb e^+ e^-$, but it is
much smaller.)

In Ref.~\cite{RKRD}, it was pointed out that, assuming the scale of NP
is much larger than the weak scale, the operator of
Eq.~(\ref{GGLoperator}) should be made invariant under the full
$SU(3)_C \times SU(2)_L \times U(1)_Y$ gauge group. One way to do this
is to write the NP operator as
\bea
{\cal O}_{NP} &=& G_2 ({\bar Q}'_L \gamma_\mu \sigma^I Q'_L) ({\bar L}'_L \gamma^\mu \sigma^I L'_L) \nn\\
&=& G_2 \left[
2 ({\bar Q}'^{i}_L \gamma_\mu Q'^{j}_L) ({\bar L}'^{j}_L \gamma^\mu L'^{i}_L)
- ({\bar Q}'_L \gamma_\mu Q'_L) ({\bar L}'_L \gamma^\mu L'_L) \right] ~,
\eea
where $G_2$ is $O(1)/\Lambda_{NP}^2$. Here $Q' \equiv (t',b')^T$ and
$L' \equiv (\nu'_\tau,\tau')^T$. The key point is that ${\cal O}_{NP}$
contains both neutral-current (NC) and charged-current (CC)
interactions. The NC and CC pieces can be used to respectively explain
the $R_K$ and $R_{D^{(*)}}$ puzzles. One NP model that contains the
above operator involves vector leptoquarks \cite{vectorLQ}. Another
assumes the addition of a set of massive vector bosons that transform
as an $SU(2)_L$ triplet, and that are coupled to both quark and lepton
currents \cite{RDRK_Isidori}. It is this second NP model that is of
interest for the $W \to \tau\nu$ puzzle.

In Ref.~\cite{HeavyVecTrip}, a formalism was presented for adding to
the SM a real spin-1 isospin-triplet $V_\mu^a$ ($a=1,2,3$) with
vanishing hypercharge. It describes heavy vector particles, one
charged ($W'$) and one neutral ($Z'$), that couple to the SM
left-handed fermionic currents. This was adapted in
Ref.~\cite{RDRK_Isidori} to the specific case where the $V$ couples
principally to the third-generation fermions. The simplified
Lagrangian is given by
\beq
{\cal L}_V = - \frac14 D_{[ \mu} V^a_{\nu ]} D^{[ \mu} V^{\nu ] a}+ \frac12 m_V^2 V_\mu^a V^{\mu a} \nn\\
+ i g_H V_\mu^a (H^\dagger T^a \overset\leftrightarrow{D}^\mu H) + V_\mu^a J^{\mu a} ~,
\label{VLagrangian}
\eeq
where $T^a = \sigma^a/2$, $D_{[ \mu} V^a_{\nu ]} \equiv D_{\mu} V^a_{\nu} -
D_{\nu} V^a_{\mu}$ with $D_{\mu} V^a_{\nu} = \partial_{\mu} V^a_{\nu} +
g \epsilon^{abc} W^b_\mu V^c_\nu$, and
\beq
J^{\mu a} = g_q \lambda_{ij}^q \left( {\bar Q}^{\prime i}_L \gamma^\mu T^a Q^{\prime j}_L \right)
+ g_l \lambda_{ij}^l \left( {\bar L}^{\prime i}_L \gamma^\mu T^a L^{\prime j}_L \right) ~.
\eeq
Here $\lambda_{ij}^{q,l}$ are Hermitian flavor matrices and
$\lambda_{33}^q = \lambda_{33}^l = 1$. In Ref.~\cite{RDRK_Isidori} it
was shown that tree-level $Z'$ and $W'$ exchange can respectively
explain the $R_K$ and $R_{D^{(*)}}$ puzzles.

It is emphasized in Ref.~\cite{HeavyVecTrip} that the $V_\mu^a$ fields
in Eq.~(\ref{VLagrangian}) are not the mass eigenstates as they mix
with the $W_\mu^a$ after electroweak symmetry breaking. In particular,
the physical $W$ mass eigenstate is
\beq
(W^\pm)_{phys} = W^\pm \cos\theta_C + V^\pm \sin\theta_C ~,
\eeq
where $\theta_C$ is the charged-current mixing angle. Naively, this
angle could be as large as $O(m_W/m_V)$, which equals 0.08 for $m_V =
1$~TeV.  In the presence of such a mixing, the \Wtaunu\
coupling is given by
\beq
g ( \cos\theta_C + (g_l/g) \lambda^l_{33} \sin\theta_C ) ~.
\eeq
The experimental measurement of the \Wtaunu\ coupling
could therefore be reproduced if the expression in parentheses equals
1.033. Given that $\lambda^l_{33} = 1$, this could happen if, for
example,
\beq
g_l = g ~~,~~~~ \theta_C = 0.034 ~.
\label{Wmixsolution}
\eeq

On the face of things, this appears to be possible. However,
constraints from the neutral-current sector must be taken into
account. In the presence of mixing, the physical $Z$ mass eigenstate
is given by
\beq
(Z^0)_{phys} = Z^0 \cos\theta_N + V^0 \sin\theta_N ~.
\eeq
The key point \cite{HeavyVecTrip} is that, for small mixing angles,
\beq
\theta_C \simeq \frac{M_W}{M_Z} \theta_N ~.
\eeq
Thus, constraints on $\theta_N$ lead directly to constraints on
$\theta_C$. And $\theta_N$ can be bounded by the data on $Z$ decays.
For example, consider $Z \to \tau^+ \tau^-$. The $Z \to \ell^+ \ell^-$
data are \cite{pdg}
\bea
{\cal B}(Z \to e^+ e^-) & = & (3.363 \pm 0.004) \% ~, \nn\\
{\cal B}(Z \to \mu^+ \mu^-) & = & (3.366 \pm 0.007) \% ~, \nn\\
{\cal B}(Z \to \tau^+ \tau^-) & = & (3.370 \pm 0.008) \% ~,
\label{Zlldecays}
\eea
leading to
\beq
\frac{ 2 {\cal B}(Z \to \tau^+ \tau^-) }{ {\cal B}(Z \to e^+ e^-) + {\cal B}(Z \to \mu^+ \mu^-) } = 1.0016 \pm 0.0027 ~.
\label{Zllrat}
\eeq
Theoretically, we have
\beq
\frac{ 2 {\cal B}(Z \to \tau^+ \tau^-) }{ {\cal B}(Z \to e^+ e^-) + {\cal B}(Z \to \mu^+ \mu^-) } =
\frac{ (a_{\tau_L}^Z)^2 + (a_{\tau_R}^Z)^2 }{(a_{\ell_L}^Z)^2 + (a_{\ell_R}^Z)^2 } ~,
\eeq
where $a_f^Z = I_{3L} - Q_{em} \sin^2 \theta_W$ is the $Z f {\bar f}$ coupling.
In the SM, the $Z \ell^+ \ell^-$ couplings are given by
\beq
a_{\ell_L}^Z = -\frac12 + \sin^2 \theta_W ~,~~ a_{\ell_R}^Z = \sin^2 \theta_W ~.
\label{SMZllcoup}
\eeq
In the presence of a $Z^0$-$V^0$ mixing, the coupling of the $Z^0$ to
left-handed $\tau$'s is modified:
\beq
a_{\tau_L}^Z = \left( -\frac12 + \sin^2 \theta_W \right) \cos\theta_N
+ \left( g_l/g\right) \cos\theta_W \lambda^l_{33} \sin\theta_N ~.
\eeq
($a_{\tau_R}^Z$ is unchanged from the SM.) Taking $\lambda^l_{33} = 1$,
$\sin^2 \theta_W = 0.231$, and $g_l = g$, this yields
\beq
-0.0026 \le \theta_N \le 0.0017 ~~(3\sigma) ~.
\eeq
This corresponds to the constraint $\theta_C < 0.0015$, which rules
out the solution of Eq.~(\ref{Wmixsolution}).

We note in passing that a similar result can be found by considering
$Z \to \nu_\tau {\bar\nu}_\tau$ decays.  In the SM,
\beq
\frac{{\cal B}(Z \to \nu_e {\bar\nu}_e)}{{\cal B}(Z \to e^+ e^-)} =
\frac{\left( \frac12 \right)^2 }{ \left( -\frac12 + \sin^2 \theta_W \right)^2 + \left( \sin^2 \theta_W \right)^2 }  ~,
\label{Znunubegin}
\eeq
so that, using Eq.~(\ref{Zlldecays}),
\beq
{\cal B}(Z \to \nu_e {\bar\nu}_e) = (6.687 \pm 0.008) \% ~.
\eeq
The SM therefore predicts that
\beq
{\cal B}(Z \to {\rm invisible}) = 3 {\cal B}(Z \to \nu_e {\bar\nu}_e) = (20.062 \pm 0.024) \% ~.
\eeq
Experimentally, we have \cite{pdg}
\beq
{\cal B}(Z \to {\rm invisible}) = (20.0 \pm 0.06) \% ~.
\eeq
As above, ${\cal B}(Z \to f {\bar f})$ is proportional to $(a_{f_L}^Z)^2 + (a_{f_R}^Z)^2$.
In the SM, the $Z \nu_\ell {\bar\nu}_\ell$ couplings are given by
\beq
a_{\nu_{\ell L}}^Z = \frac12 ~,~~ a_{\nu_{\ell R}}^Z = 0 ~.
\eeq
In the presence of $Z^0$-$V^0$ mixing, we have
\beq
a_{\nu_{\tau L}}^Z = \frac12 \cos\theta_N + \left( g_l/g\right) \cos\theta_W \lambda^l_{33} \sin\theta_N ~~,~~~~ a_{\nu_{\ell R}}^Z = 0 ~,
\eeq
with
\bea
\frac{{\cal B}(Z \to {\rm invisible})_{NP}}{{\cal B}(Z \to {\rm invisible})_{\SM}} & = &
\frac{ (a_{\nu_{\tau L}}^Z)^2 + 2 \left( \frac12 \right)^2} { 3 \left( \frac12 \right)^2} \nn\\
& = & \frac{20.0 \pm 0.06}{20.061 \pm 0.014} = 0.997 \pm .003 ~.
\label{Znunuend}
\eea
Taking $\lambda^l_{33} = 1$, $\sin^2 \theta_W = 0.231$, and $g_l = g$,
we obtain
\beq
-0.0093 \le \theta_N \le 0.0046 ~~(3\sigma) ~.
\eeq
This corresponds to the constraint $\theta_C < 0.004$. This is weaker
than that from $Z \to \tau^+ \tau^-$, but it still rules out the
solution of Eq.~(\ref{Wmixsolution}).

We therefore conclude that the $W \to \tau\nu$ puzzle cannot be
explained by $W$-$W'$ mixing.

\section{Mixing with Isospin-triplet Leptons}

In this section we consider the mixing of $\tau_{L,R}$ and $\nu_{\tau
  L}$ with isospin-triplet leptons. Such exotic leptons were examined
in Ref.~\cite{Delgado}, and were allowed to mix with all three flavors
of SM leptons. This then generates flavor-changing neutral-current
processes (FCNCs) such as $\mu \to e\gamma$, $\tau\to\mu\mu\mu$,
etc. Reference~\cite{Delgado} focused specifically on FCNCs, as well as on
the phenomenology of the exotic leptons.

In the present paper, the isospin-triplet leptons are allowed to mix
with only one flavor of SM leptons, $\tau$ and $\nu_\tau$, so
that FCNCs are not generated. Thus, only flavor-conserving processes
(such as $W \to \tau\nu$) are affected. Now, if the new leptons with
which $\tau_L$ and $\nu_{\tau L}$ mix were singlets under $SU(2)_L$,
the $W$-$\tau$-$\nu$ coupling would be reduced (by the cosine of the
mixing angle) \cite{LL}.  However, as we show below, if the exotic
leptons are isospin triplets, this coupling can be increased.

We consider two types of isospin triplets:
\beq
L_{L,R} \equiv
\left(
\begin{array}{c}
L^+~ \\ L^0~ \\ L^-
\end{array}
\right)_{L,R}
~~,~~~~
L'_{L,R} \equiv
\left(
\begin{array}{c}
L^{\prime 0}~ \\ L^{\prime -}~ \\ L^{\prime --}
\end{array}
\right)_{L,R}
~.
\eeq
$L$ has hypercharge $Y=0$ and is Majorana; $L'$ has hypercharge $Y=-2$
and is Dirac. Both are vector fermions, i.e., their $L$ and $R$
chiralities are both isospin triplets. As shown in the Appendix
[Eq.~(\ref{CCisotrip})], since $L^{(\prime)}$ is an isotriplet, the
charged-current interactions between $L^{(\prime) 0}$ and $L^{(\prime)
  -}$ take the form
\beq
g \left[ {\bar L}^{(\prime) 0} \gamma^\mu W_\mu^+  L^{(\prime) -} + {\bar L}^{(\prime) -} \gamma^\mu W_\mu^-  L^{(\prime) 0} \right] ~.
\label{CCisotrip}
\eeq
Compare this to the SM charged-current interaction terms:
\beq
\frac{1}{\s} g \left[ {\bar\nu}_\tau \gamma^\mu W_\mu^+ \gamma_L \tau^- + {\bar\tau}^- \gamma^\mu W_\mu^- \gamma_L \nu_\tau \right] ~.
\eeq
It is the different coefficients -- $1/\sqrt{2}$ for isospin doublets,
1 for isospin triplets -- that has the potential to produce an
increased $W$-$\tau$-$\nu$ coupling.

The basic idea is as follows. The SM fermions are
\beq
E_L \equiv
\left(
\begin{array}{c}
\nu_\tau \\ \tau^-
\end{array}
\right)_L
~~,~~~~
\tau^-_R ~.
\eeq
Both $L_L$ and $L'_L$ have components with $Q_{em} = -1$ ($L^{(\prime)
  -}_L$) and $Q_{em} = 0$ ($L^{(\prime) 0}_L$). Suppose the $\tau^-_L$
and $\nu_{\tau L}$ mix with these. We then have
\bea
\label{LHtaumixing}
(\tau^-_L)_{phys} &=& \tau^-_L \cos\theta_L^\tau + L^{(\prime) -}_L \sin\theta_L^\tau ~, \\
(\nu_{\tau L})_{phys} &=& \nu_{\tau L} \cos\theta_L^\nu + L^{(\prime) 0}_L \sin\theta_L^\nu ~.
\label{LHnumixing}
\eea
In the presence of mixing, the charged current between the physical
$\tau^-_L$ and $\nu_{\tau L}$ has two pieces: $\tau^-_L$-$\nu_{\tau
  L}$ (isospin doublet) and $L^{(\prime) -}_L$-$L^{(\prime) 0}_L$
(isospin triplet).  The strength of the $W$-$\tau$-$\nu$ coupling
therefore changes:
\beq
\left. \frac{1}{\s} \right\vert_{\SM} \to \frac{1}{\s} (\cos\theta_L^\tau \cos\theta_L^\nu + \s \sin\theta_L^\tau \sin\theta_L^\nu) ~.
\label{correction}
\eeq
As was the case with $W$-$W'$ mixing, the experimental measurement of
the \Wtaunu\ coupling could be reproduced if the
expression in parentheses equals 1.033. This could happen if, for
example,
\beq
\theta_L^\tau = \theta_L^\nu ~~,~~~~ \sin\theta_L^\tau = 0.28 ~.
\label{LHmixangle}
\eeq

One immediate question is: theoretically, can such large mixing be
obtained? In the case of mixing with isotriplet leptons, the answer is
yes. Because both $E_L $ and the SM Higgs are doublets under
$SU(2)_L$, and because ${\bf 2} \otimes {\bf 2} = {\bf 1} \oplus {\bf
  3}$, one can write dimension-4 operators that involve $E_L$, $H$,
and $L^{(\prime)}$.  When the Higgs acquires a vacuum expectation
value (VEV) $v/\s$, it generates a mass term mixing $E_L$ and
$L^{(\prime)}$. This mass term $m$ is naturally of $O(v)$. Assuming
the exotic leptons have masses $M \simeq O($1 TeV), the mixing angle
will be $O(m/M)$. This is in the right ballpark of the above angle.

On the other hand, this would not work if $\tau_{L,R}$ and $\nu_{\tau
  L}$ mix with exotic leptons of higher isospin. In this case,
higher-dimension operators involving more than one Higgs field are
required. These are suppressed by powers of the NP mass scale, and so
the mixing angles will be correspondingly reduced.

We therefore see that the mixing of $\tau_{L,R}$ and $\nu_{\tau L}$
with exotic isotriplet leptons has the potential to explain the $W \to
\tau\nu$ puzzle. But this raises further questions. Is such a mixing
consistent with other experimental constraints? If not, are there any
mixing scenarios, even fine-tuned, in which this can be made to work?
To investigate these questions, we consider four different models
involving the mixing of $\tau_{L,R}$ and $\nu_{\tau L}$ with
isotriplet leptons: \\
\null\qquad (i) mixing with $L'$ alone, \\
\null\qquad (ii) mixing with $L$ alone, \\
\null\qquad (iii) mixing with $L$ and $L'$, \\
\null\qquad (iv) mixing with $L$, $L'$ and $\nu_{\tau R}$.

For a given model to pass all the experimental tests, it must (1) give
the correct value of $m_\nu$, (2) reproduce the measured value of the
\Wtaunu\ coupling, and (3) satisfy the constraints from $Z \to \tau^+
\tau^-$ and $Z \to \nu_\tau {\bar\nu}_\tau$. Regarding test (1), we
know that $\nu_\tau$ has a tiny mass, and there are neutrino
oscillations. However, as can be seen in the Appendix, the nonzero
entries in the neutrino mass matrices are all $O(v)$ or $O($1 TeV). As
such, there is no way of implementing a seesaw mechanism, which
requires a mass term of $O(10^{15}$ GeV). For this reason we require
only that a model contain a massless neutrino in order to pass test
(1).  If a possible solution to the $W \to \tau\nu$ puzzle is found,
we can then try to explain neutrino masses and oscillations by
allowing all three neutrinos to mix and incorporating some sort of
seesaw mechanism\footnote{In principle, the \Wtaunu\ puzzle could be
  connected to the induced nonunitarity of the 
  Pontecorvo-Maki-Nakagawa-Sakata (PMNS) matrix that describes the
  mixing between massive neutrino species. In Ref.~\cite{numix}, a
  global fit to precision data from varying energy ranges was
  performed in the Minimal Unitarity Violation scheme. It was shown
  that the data do indeed prefer a small amount of non-unitarity in
  leptonic mixing. However, the best-fit point was found to slightly
  worsen the \Wtaunu\ puzzle. In the present work, we ignore neutrino
  mixing, and hence do not consider the complications arising from a
  non-unitary PMNS matrix.}. However, for now we are content to focus
on massless neutrinos.

Models (i)-(iv) are analyzed in detail in the Appendix. In all cases,
we first examine the mass matrix of the neutral leptons, to see if the
model passes test (1), i.e., if it predicts a tiny mass or $m=0$ for
$\nu_\tau$. We find that models (ii) and (iii) fail this test.
However, models (i) and (iv) do contain $\nu_\tau$ with $m=0$. For
these models, we express $(\nu_{\tau L})_{phys}$ and
$(\tau^-_{L,R})_{phys}$ in terms of the gauge eigenstates. This allows
us to move on to tests (2) and (3).

Consider first model (i). Here the $\tau^-_L$ mixes with $L^{\prime
  -}_L$ and the $\nu_{\tau L}$ mixes with $L^{\prime 0}_L$ as in
Eqs.~(\ref{LHtaumixing}) and (\ref{LHnumixing}). However, according to
Eqs.~(\ref{model(i)numix}) and (\ref{model(i)taumix}) in the Appendix,
the mixing angles obey
\beq
\sin\theta_L^\tau \simeq \frac{m'_2}{\sqrt{2} M'} ~~,~~~~
\sin\theta_L^\nu \simeq - \frac{m'_2}{M'} ~.
\label{signdiff}
\eeq
That is, they are of opposite sign\footnote{This sign difference is
  due to the opposite signs of the $m'_2$ entries in the neutral- and
  charged-lepton mass matrices [Eqs.~(\ref{nuMmatrix}) and
    (\ref{lMmatrix})]. And this is in turn due to fact that the 12 and
  22 elements of the $2\times 2$ representation of $L'$
  [Eq.~(\ref{L2x2})], which contribute to the mass matrices, are of
  opposite sign.}. Since the correction to the \Wtaunu\
coupling is proportional to $\sin\theta_L^\tau
\sin\theta_L^\nu$, mixing actually has the effect of {\it
  decreasing} the coupling. Model (i) thus fails test (2).

For completeness, how does model (i) fare with test (3)? First,
consider $Z \to \tau^+ \tau^-$ [see
  Eqs.~(\ref{Zlldecays})-(\ref{SMZllcoup})]. In the
presence of mixing, the $I_{3L}$ of the physical $\tau^-_L$ is
\beq
\langle (\tau^-_L)_{phys} \vert T_3 \vert (\tau^-_L)_{phys} \rangle =
-\frac12 \cos^2\theta_L^\tau + 0 \cdot \sin^2\theta_L^\tau = -\frac12 \left( 1 - \sin^2\theta_L^\tau \right) ~.
\eeq
[Even with mixing, $(\tau^-_R)_{phys}$ still has $I_{3L} = 0$, as in
the SM.] This implies [see Eqs.~(\ref{Zllrat})-(\ref{SMZllcoup})]
\beq
\frac{ 2 {\cal B}(Z \to \tau^+ \tau^-) }{ {\cal B}(Z \to e^+ e^-) + {\cal B}(Z \to \mu^+ \mu^-) } =
\frac{ (-\frac12 \left( 1 - \sin^2\theta_L^\tau \right) + \sin^2 \theta_W)^2 + (\sin^2 \theta_W)^2 }
             { (-\frac12 + \sin^2 \theta_W)^2 + (\sin^2 \theta_W)^2 } = 1.0016 \pm 0.0027 ~,
\eeq
leading to
\beq
|\sin\theta_L^\tau| \le  0.055 ~~(3\sigma) ~.
\label{taulimit}
\eeq

Second, consider $Z \to \nu_\tau {\bar\nu}_\tau$ [see
  Eqs.~(\ref{Znunubegin})-(\ref{Znunuend})]. In the presence of
mixing, the $I_{3L}$ of the physical $\nu_{\tau L}$ is
\beq
\langle (\nu_{\tau L})_{phys} \vert T_3 \vert (\nu_{\tau L})_{phys} \rangle =
\frac12 \cos^2\theta_L^\nu + 1 \cdot \sin^2\theta_L^\nu
= \frac12 \left( 1 + \sin^2\theta_L^\nu \right) ~.
\eeq
This modified $I_{3L}$ will affect the $Z \nu_\tau {\bar\nu}_\tau$
coupling, so that [see Eq.~(\ref{Znunuend})]
\beq
\frac{{\cal B}(Z \to {\rm invisible})_{NP}}{{\cal B}(Z \to {\rm invisible})_{\SM}} =
\frac{ \left( \frac12 (1 + \sin^2\theta_L^\nu) \right)^2 + 2 \left( \frac12 \right)^2}
{ 3 \left( \frac12 \right)^2} = 0.997 \pm .003 ~.
\eeq
This implies that
\beq
|\sin\theta_L^\nu| \le  0.099 ~~(3\sigma) ~.
\label{nulimit}
\eeq

Combining Eqs.~(\ref{taulimit}) and (\ref{nulimit}), we have
\beq
\cos\theta_L^\tau \cos\theta_L^\nu + \s \sin\theta_L^\tau \sin\theta_L^\nu < 1.002 ~.
\eeq
Following Eq.~(\ref{correction}), it was noted that this quantity
should equal 1.033 to explain the $W \to \tau\nu$ puzzle. This is clearly
not satisfied by the above equation. Thus, even if the sign difference
of Eq.~(\ref{signdiff}) had not been present, model (i) could not have
passed test (2). Conversely, taking values for the mixing angles large
enough to explain the $W \to \tau\nu$ puzzle would have resulted in
failing test (3). The bottom line is that model (i) does not pass the
experimental tests.

We now turn to model (iv). Here the mixing is much more
complicated. The expressions for $(\nu_{\tau L})_{phys}$ and
$(\tau^-_{L,R})_{phys}$ are given in Eqs.~(\ref{nuLmodel(iv)}) and~
(\ref{tauLmodel(iv)}), and are repeated for convenience below:
\bea
(\tau^-_L)_{phys} & = & a_{L\tau} \tau^-_L + c_{L\tau} L^-_L + d_{L\tau} L^{\prime -}_L ~, \nn\\
(\nu_{\tau L})_{phys} & = & a_\nu \nu_{\tau L} + b_\nu \nu_{\tau R}^c + c_\nu {L^0_R}^c + d_\nu L^{\prime 0}_L + e_\nu {L^{\prime 0}_R}^c ~.
\eea
Expressions for the coefficients are given in Eqs.~(\ref{nucoeffs})
and (\ref{taucoeffs}), in terms of the various mass parameters that
appear in the relevant mixing matrices. In order to obtain a massless
$\nu_\tau$, we require
\bea
\frac{m_D^2}{m_S^2} &=& \eta \frac{m_2^2}{2 M^2}~,
\label{zero}
\eea
where $\eta = - M / m_S$.

We begin by considering effects of this mixing on the neutral-current
sector. In the presence of mixing, the $I_{3L}$ of the physical
$\tau^-_L$ is
\bea
\langle (\tau^-_L)_{phys} \vert T_3 \vert (\tau^-_L)_{phys} \rangle  &=&
-\frac12  + \frac12 ( 1-(a_{L}^{\tau} )^2) - (c_{L}^{\tau})^2 ~, \nl
&\approx& -\frac12   \left [1+
\frac{ m_2^2}{{M}^2 } - \frac{ {m'_2}^2}{{2M'}^2 }   \right]~,
\label{tau}
\eea
while that of the physical $\nu_{\tau L}$ is
\bea
\langle (\nu_{\tau L})_{phys} \vert T_3 \vert (\nu_{\tau L})_{phys} \rangle
&=& \frac12  \left[ (a_{L}^{\nu})^2 +
2 \left ((d_{L}^{\nu})^2 -  (e_{L}^{\nu})^2 \right) \right] \nl
&\approx& \frac12 \left [1 - (1+ \eta) \dfrac{m^2_2}{2M^2}
+ \frac{{m'_2}^2}{{M'}^2}\right]~.
\label{nu}
\eea
In both cases above, we have expanded the expressions for the
coefficients, neglecting $m_\tau$ and keeping terms to leading order
in the mixing parameters $m^2_2/M^2$ and $m'^2_2/M'^2$. Now, we saw in
the study of model (i) that the constraints from the decays $Z \to
\tau^+\tau^-$ and $Z\to \nu_\tau {\bar\nu}_\tau$ are quite severe. To
evade these constraints, the mass parameters in model (iv) must be
such that the values of $I_{3L}$ of both $(\tau^-_L)_{phys}$ and
$(\nu_{\tau L})_{phys}$ are unchanged from their SM values, i.e.,
$I_{3L} = -1/2$ ($(\tau^-_L)_{phys}$) and $I_{3L} = 1/2$ ($(\nu_{\tau
  L})_{phys}$).  Equations ~(\ref{tau}) and ~(\ref{nu}) then imply that $2
m^2_2/M^2 = m'^2_2/M'^2$ and $\eta = 3$.

Now, the \Wtaunu\ coupling in this model is proportional to
\bea
K &=& a_{L\tau} a_\nu + \s~(c_{L\tau} c_\nu + d_{L\tau} d_\nu) \nn\\
&=& \dfrac{1 + \dfrac{m^2_2}{M^2 - m^2_\tau} - \dfrac
{{m'_2}^2}{{M'}^2 - m^2_\tau}}{\sqrt{\(1 + \dfrac{m^2_D}{m^2_S}
+ \dfrac{m^2_2}{2M^2} + \dfrac{{m'_2}^2}{{M'}^2}\)\(1 + \dfrac
{m^2_2 M^2}{(M^2 - m^2_\tau)^2} + \dfrac{{m'_2}^2 {M'}^2}{2({M'}^2 -
m^2_\tau)^2}\)}} ~, \\
&\approx& 1  - \dfrac{7{m'_2}^2}{4{M'}^2} +  \frac{(1- \eta)}{4}
\dfrac{m^2_2}{M^2 } ~,
\eea
where we have once again neglected $m_\tau$ and kept only the
leading-order terms in $m^2_2/M^2$ and $m'^2_2/M'^2$ in the
expansion. Above, it was noted that $\eta = 3$ is required to evade
the constraints from $Z \to \tau^+\tau^-$ and $Z\to \nu_\tau
{\bar\nu}_\tau$. However, if $ \eta \ge 1$, we have $K < 1$, so that,
as was the case with model (i), mixing has the effect of reducing the
\Wtaunu\ coupling. Thus, although the model passes test (3), it now
fails test (2). We therefore conclude that the $W \to \tau\nu$ puzzle
cannot be explained by allowing $\tau_{L,R}$ and $\nu_{\tau L}$ to mix
with isospin-triplet leptons.

\section{Conclusions}

At present, there are several measurements of $B$ decays that exhibit
discrepancies with the predictions of the SM at the level of $2\sigma$
or greater. These hints of new physics have been taken very seriously
-- there has been a flurry of theoretical activity looking for NP
explanations of the various $B$-decay results. Another decay that has
a similar disagreement with the SM is $W \to \tau\nu$. According to
the Particle Data Group, there is a $2.3\sigma$ disagreement between
${\cal B}(W^+ \to \tau^+ \nu_\tau)$ and ${\cal B}(W^+ \to \ell^+
\nu_\ell)$ ($\ell = e,\mu$). However, for some reason -- perhaps
because $\tau^- \to e^- \nu_\tau {\bar\nu}_e$ does not exhibit a
similar discrepancy with the SM -- little attention has been paid to
this result. In the present paper, we search for a NP explanation of
the $W \to \tau\nu$ puzzle.

The obvious conclusion to be drawn from the experimental measurement
is that the \Wtaunu\ coupling has been increased due to the presence
of NP. Because the process is rather simple -- an on-shell $W$
decaying to $\tau\nu$ -- there are only two possible ways NP can
enter. Either the $W$ mixes with a $W'$, or the $\tau_L$ and the
$\nu_{\tau L}$ mix with exotic leptons. We consider both
possibilities.

First, we assume that a $W'$ boson exists that couples preferentially
to the third generation. $W$-$W'$ mixing could then lead to an
increased \Wtaunu\ coupling. The problem is that such a $W'$ also
comes with a neutral partner, a $Z'$, that mixes with the SM $Z$. Now,
the amounts of $W$-$W'$ and $Z$-$Z'$ mixings are related. And $Z$-$Z'$
mixing is strongly constrained by the experimental measurements of $Z
\to \tau^+\tau^-$ and $Z\to \nu_\tau {\bar\nu}_\tau$. The upshot is
that, when the constraints from $Z$ decays are taken into account, the
allowed $W$-$W'$ mixing is too small to produce the necessary increase
in the \Wtaunu\ coupling.

Second, we allow $\tau_{L,R}$ and $\nu_{\tau L}$ to mix with
isospin-triplet leptons. Such mixing can potentially lead to an
increased \Wtaunu\ coupling. We take two isospin-triplet leptons, one
with hypercharge $Y=0$, the other with $Y=-2$, and consider a variety
of mixing scenarios. For a given scenario to succeed, it must (1) give
the correct value of $m_{\nu_\tau}$ ($m=0$ is allowed), (2) reproduce
the measured value of the \Wtaunu\ coupling, and (3) satisfy the
constraints from $Z \to \tau^+ \tau^-$ and $Z \to \nu_\tau
{\bar\nu}_\tau$.  Unfortunately, all the mixing scenarios fail at
least one of these tests.

Because we are unable to find a NP explanation of the $W \to \tau\nu$
puzzle, we are forced to conclude that it is almost certainly just a
statistical fluctuation.

\bigskip
\noindent
{\bf Acknowledgments}: We would like to thank P. Langacker for the helpful
communications, and A. Greljo, G. Isidori and D. Marzocca for answers
to questions about Ref.~\cite{RDRK_Isidori}. This work was financially
supported by NSERC of Canada (B.~B., D.~L.), and by the National Science
Foundation (A.~D.) under Grant No.\ NSF PHY-1414345.

\section*{Appendix}
\appendix

We consider two types of isospin triplets:
\beq
L_{L,R} \equiv
\left(
\begin{array}{c}
L^+~ \\ L^0~ \\ L^-
\end{array}
\right)_{L,R}
~~,~~~~
L'_{L,R} \equiv
\left(
\begin{array}{c}
L^{\prime 0}~ \\ L^{\prime -}~ \\ L^{\prime --}
\end{array}
\right)_{L,R}
~.
\eeq
$L$ and $L'$ have hypercharge $Y=0$ and $Y=-2$, respectively.  Both
are vector fermions, i.e., their $L$ and $R$ chiralities are both
isospin triplets. For this reason, they may have direct mass terms.
However, $L$ is Majorana, while $L'$ is Dirac, which means that the
forms of the mass terms and mass matrices are different. The kinetic
and direct mass terms for $L$ and $L'$ are
\bea
{\cal L}_{\rm kin} & = & \frac{1}{2}{\bar L}_a i \gamma^\mu D^{ab}_\mu L_b
- \frac{M}{2}({\bar L}^c_aL_a + {\bar L}_aL^c_a) ~, \nn\\
{\cal L}'_{\rm kin} & = & {\bar L}'_a i \gamma^\mu D^{ab}_\mu L'_b - M'{\bar L}'_aL'_a~,
\label{kinetic}
\eea
with
\beq
D_\mu = \partial_\mu - i \frac{g}{\s} (W_\mu^+ T^+ + W_\mu^- T^-)
- i \frac{g}{\cos\theta_W} Z_\mu (T^3 - Q_{em} \sin^2 \theta_W) -i e A_\mu Q_{em} ~.
\eeq
In the above, $a, b = 1, 2, 3$ are isospin indices. Although the
covariant derivative $D_\mu$ itself is representation independent, the
form of the SU(2) generators ($T^\pm, T_3$) depends on whether the
fermion is an isodoublet or an isotriplet. In particular, the
charged-current interactions between the isotriplets $L^{(\prime) 0}$
and $L^{(\prime) -}$ are
\beq
g \left[ {\bar L}^{(\prime) 0} \gamma^\mu W_\mu^+  L^{(\prime) -} + {\bar L}^{(\prime) -} \gamma^\mu W_\mu^-  L^{(\prime) 0} \right] ~.
\label{CCisotrip}
\eeq

We examine models in which combinations of the following particles
mix:
\beq
E_L \equiv
\left(
\begin{array}{c}
\nu_\tau \\ \tau^-
\end{array}
\right)_L
~~,~~~~
\tau^-_R
~~,~~~~
\nu_{\tau R}
~~,~~~~
L_{L,R}
~~,~~~~
L'_{L,R}
~.
\eeq
The direct mass terms for $L$ and $L'$ are shown above
[Eq.~(\ref{kinetic})]; the mass terms for $E_L$, $\tau_R$ and
$\nu_{\tau R}$ are given by
\beq
-\lambda_1 {\bar E}_L H \tau_R
- \lambda_D {\bar E}_L \tilde{H}^\dag \nu_{\tau R}
- \frac{m_S}{2}\bar{\nu_{\tau R}^c}\nu_{\tau R}^{ }  + {\rm H.c.}~.
\eeq
Here $H=\(\bma\phi^+ & \phi^0\ema\)^T$ and $\tilde{H} = \(\bma \phi^0
& -\phi^+\ema\)$.  When the Higgs acquires a VEV $v/\s$, the $\tau$
and $\nu_\tau$ obtain Dirac masses $m_1 \equiv \lambda_1 v/\s$ and
$m_D \equiv \lambda_D v/\s$, respectively. The $\nu_{\tau R}$ also has
a Majorana mass $m_S/2$. $m_D$ is $O(v)$, while the size of $m_S$ is
unspecified. Mixing between $E_L$ and $L^{(\prime)}_R$ is generated by
the Yukawa terms
\beq
- \lambda_2 {\bar E}_L L_R\tilde{H}^\dag
- \lambda'_2 {\bar E}_L L'_R H  + {\rm H.c.}~ ~.
\label{Yukawa}
\eeq
The mixing terms $m_2 \equiv \lambda_2 v/2$ and $m'_2 \equiv
\lambda'_2 v/2$ are both $O(v)$. In the above equation, $L$ and $L'$
are expressed as $2\times 2$ matrices:
\beq
L = \frac{1}{2}\(\bma L^0 & \s L^+ \\ \s L^- & - L^0 \ema\)
~~,~~~~
L' = \frac{1}{2}\(\bma L^{\prime -} & \s L^{\prime 0} \\ \s L^{\prime --} & - L^{\prime -} \ema\) ~.
\label{L2x2}
\eeq

Below we examine four different models: $\tau_{L,R}$ and $\nu_{\tau
  L}$ mixing with (i) $L'$ alone, (ii) $L$ alone, (iii) $L$ and $L'$,
and (iv) $L$, $L'$ and $\nu_{\tau R}$. If a given model does not contain a
neutrino with a tiny mass ($m=0$ is accepted), it is excluded from
further consideration. If it passes this test, we find the
eigenvectors corresponding to $(\tau_{L,R})_{phys}$ and $(\nu_{\tau
  L})_{phys}$.

\subsection{\boldmath $\tau_{L,R}$ and $\nu_{\tau L}$ mixing with $L'$ alone}

For the neutral leptons the mass terms translate to
\beq
-\left(
\begin{array}{cc}
{\bar\nu}_\tau & {\bar L}^{\prime 0}
\end{array}
\right)_L
{\cal M}_{\nu L'}
\left(
\begin{array}{c}
\nu_{\tau L}^c \\
L^{\prime 0}_R
\end{array}
\right)
+ {\rm H.c.}
~~,~~~~
{\rm with} ~~~~
{\cal M}_{\nu L'} =
\left(
\begin{array}{cc}
0 & m'_2 \\
0  & M' \\
\end{array}
\right) ~,
\label{nuMmatrix}
\eeq
while for the charged leptons the mass terms take the form
\beq
-\left(
\begin{array}{cc}
{\bar\tau}^- & {\bar L}^{\prime -}
\end{array}
\right)_L
{\cal M}_{\tau L'}
\left(
\begin{array}{c}
\tau^- \\
L^{\prime -}
\end{array}
\right)_R  + {\rm H.c.} ~~,~~~~
{\rm with} ~~~~
{\cal M}_{\tau L'} =
\left(
\begin{array}{cc}
m_1 & -m'_2/\s \\
0 & M' \\
\end{array}
\right) ~.
\label{lMmatrix}
\eeq
Recall that $m'_2$ and $M'$ are, respectively, $O(v)$ and $O($1 TeV).
Note that, because of gauge invariance, there is no mass term relating
$L^{\prime -}_L$ and $\tau^-_R$.

The states in Eqs.~(\ref{nuMmatrix}) and (\ref{lMmatrix}) are
defined in the gauge basis. To transform to the mass basis one
applies the unitary transformations $U_L$ and $U_R$ on the left-handed
and right-handed states, respectively. $U_L$ ($U_R$) diagonalizes
${\cal M} {\cal M}^\dagger$ (${\cal M}^\dagger {\cal M}$). In the
present case, since the mass matrices are real, the transformation
matrices are orthogonal, $O_L$ and $O_R$. The diagonalization of the
mass matrices yields the mass eigenvalues and the decomposition of the
mass eigenstates in terms of gauge eigenstates.

For the neutral leptons, this procedure is rather simple. The mass
eigenvalues are $m=0$ and $m = \sqrt{{m'_2}^2 + {M'}^2}$. The
eigenstate that has $m=0$ is given by Eq.~(\ref{LHnumixing}), with
\beq
\sin\theta_L^\nu = -\frac{m'_2}{\sqrt{{m'_2}^2 + {M'}^2}} ~.
\label{model(i)numix}
\eeq
This is $O(v/M')$.

Turning to the charged-lepton mass matrix, we assume that the lighter
of the two eigenvalues is the physical $\tau$-lepton mass ($m_\tau$).
We find that the eigenstates with $m = m_\tau$ are given by
Eq.~(\ref{LHtaumixing}) (and its analogue for $\tau_R$), with
\bea
\sin\theta_L^\tau &=& \frac{m'_2}{\sqrt{{m'_2}^2 + 2 {M'}^2(1 - m^2_\tau/{M'}^2)^2}} ~, \nn\\
\sin\theta_R^\tau &=& \frac{\s m_1 m'_2}{\sqrt{2m^2_1{m'_2}^2 + ({m'_2}^2 + 2 {M'}^2 - 2 m^2_\tau)^2}} ~.
\label{model(i)taumix}
\eea
Note that, while $\sin\theta_L^\tau = O(v/M')$,
$\sin\theta_R^\tau$ is much smaller, $O(m_\tau v/{M'}^2)$.

\subsection{\boldmath $\tau_{L,R}$ and $\nu_{\tau L}$ mixing with $L$ alone}

For the neutral leptons, due to the Majorana nature of $L$, the mass
terms take the form
\beq
-\left(
\begin{array}{cc}
{\bar\nu}_{\tau L} & {\overline{{L^0_R}^c}}
\end{array}
\right)_L
{\cal M}_{\nu L}
\left(
\begin{array}{c}
\nu_{\tau L}^c \\
L^0_R
\end{array}
\right)  + {\rm H.c.}~~,~~~~
{\rm with} ~~~~
{\cal M}_{\nu L} =
\left(
\begin{array}{cc}
0 & m_2/2\s \\
m_2/2\s & M/2
\end{array}
\right) ~.
\eeq
The mass eigenvalues for the neutral leptons are obtained by
diagonalizing ${\cal M}_{\nu L}$. Approximately, these are $m=
-m_2^2/4M$ and $m = M/2$. However, we see immediately that this is
problematic. Given that $m_2$ is $O(v)$ and $M$ is $O($1 TeV), the
light neutrino mass $m_2^2/4M$ is orders of magnitude too large. We
therefore conclude that the model in which $\tau_{L,R}$ and $\nu_{\tau
  L}$ mix with $L$ alone is not viable.

\subsection{\boldmath $\tau_{L,R}$ and $\nu_{\tau L}$ mixing with $L$ and $L'$}

The mass term for the neutral leptons takes the form
\beq
-\(\bma
{\overline{\nu_{\tau L}}} & {\overline{{L^0_R}^c}} &
{\overline{L^{\prime 0}_L}} & {\overline{{L^{\prime 0}_R}^c}}
\ema\)
{\cal M}_{\nu LL'}\(\bma \nu_{\tau L}^c \\ L^0_R \\ {L^{\prime 0}_L}^c \\
L^{\prime 0}_R \ema\)  + {\rm H.c.} ~~,~~~~
{\rm with} ~~~~
{\cal M}_{\nu LL'} =
\(\bma 0 & m_2/2\s & 0 & m'_2/2 \\
m_2/2\s & M/2 & 0 & 0 \\
0 & 0 & 0 & M'/2 \\
m'_2/2 & 0 & M'/2 & 0 \ema \) ~.
\eeq
Now, the determinant of the diagonalized matrix -- which is just the
product of the four mass eigenvalues -- is equal to the determinant of
the above mass matrix.  However, ${\rm Det}({\cal M}_{\nu LL'}) =
m^2_2 {M'}^2/32$, which is nonzero. So this mass matrix does not yield
an $m=0$ eigenvalue.  Furthermore, since $m_2$, and $m'_2$ are $O(v)$,
while $M$ and $M'$ are $O($1 TeV), there is no possibility of a seesaw
mechanism. It is therefore not possible to generate a tiny mass for
$\nu_\tau$, which rules out the model in which $\tau_{L,R}$ and
$\nu_{\tau L}$ mix with $L$ and $L'$.

\subsection{\boldmath $\tau_{L,R}$ and $\nu_{\tau L}$ mixing with $L$, $L'$ and
  $\nu_{\tau R}$}

The mass terms for neutral leptons take the following form:
\beq
-\(\bma {\overline{\nu_{\tau L}}} & {\overline{\nu_{\tau R}^c}} & {\overline{{L^0_R}^c}} & {\overline{L^{\prime 0}_L}}
& {\overline{{L^{\prime 0}_R}^c}} \ema\)
{\cal M}_{\nu\nu LL'}\(\bma \nu_{\tau L}^c \\ \nu_{\tau R} \\ L^0_R \\ {L^{\prime 0}_L}^c \\ L^{\prime 0}_R \ema\) + {\rm H.c.} ~~,~~~~
{\rm with} ~~~~
{\cal M}_{\nu\nu LL'} =
\(\bma
0 & m_D/2 & m_2/2\s & 0 & m'_2/2 \\
m_D/2 & m_S/2 & 0 & 0 & 0 \\
m_2/2\s & 0 & M/2 & 0 & 0 \\
0 & 0 & 0 & 0 & M'/2 \\
m'_2/2 & 0 & 0 & M'/2 & 0
\ema\) ~.
\eeq
$m_D$, $m_2$, and $m'_2$ are $O(v)$, while $M$ and $M'$ are $O($1 TeV).
However, the size of $m_S$ is as yet undetermined.

If $m_S$ were $O(10^{15}$ GeV), it might be possible to generate a
tiny mass of $O(v^2/m_S)$ for $\nu_\tau$ via the seesaw mechanism.  In
order to establish whether this is possible, it is necessary to
diagonalize ${\cal M}_{\nu\nu LL'}$. However, this involves solving a
quintic equation, which cannot be done analytically. Still, we can get
some information about the mass eigenvalues as follows. First, we know
that $m_D$, $m_2$, and $m'_2$ are less than $M$ and $M'$. In the limit
where these entries are neglected, the mass eigenvalues are 0,
$m_S/2$, $M/2$, $M'/2$, and $-M'/2$, i.e., there are three intermediate
mass eigenvalues of $O($1 TeV). When $m_D$, $m_2$, and $m'_2$ are
included, the values of these eigenvalues will be modified. However,
we do not expect them to change enormously -- perhaps a multiplicative
factor of $10^{\pm 1}$ is possible. Second, the determinant of the
diagonalized matrix -- which is just the product of the five mass
eigenvalues -- is equal to ${\rm Det}({\cal M}_{\nu\nu LL'}) = {M'}^2
(2 M m^2_D + m^2_2 m_S)/64$. If $m_S$ is $O(10^{15}$ GeV), this is
$O(10^{24}~{\rm GeV}^5)$. Given that $m_{\nu_\tau} = O(v^2/m_S) \sim
10^{-11}~{\rm GeV}$, this implies that the product of the three
intermediate mass eigenvalues is $O(10^{20}~{\rm GeV}^3)$. However, as
we argued above, this is many orders of magnitude larger than what is
possible with ${\cal M}_{\nu\nu LL'}$. A seesaw mechanism can
therefore not be implemented.

It is still possible to generate a neutrino mass eigenvalue $m=0$ if
$2 M m^2_D + m^2_2 m_S = 0$. Keeping in mind the expected sizes of
$m_D$, $m_2$, and $M$, a simple choice that satisfies this condition is
$m^2_D/m^2_S = \eta m^2_2/(2 M^2)$ with $\eta = - M/m_s > 0$. This is clearly
a fine-tuned solution, but we cannot overlook any possibilities. For this
solution, we have
\beq
(\nu_{\tau L})_{phys} = a_\nu \nu_{\tau L} + b_\nu \nu_{\tau R}^c + c_\nu {L^0_R}^c + d_\nu L^{\prime 0}_L + e_\nu {L^{\prime 0}_R}^c ~.
\label{nuLmodel(iv)}
\eeq
The coefficients are found as follows. Defining $V_\nu \equiv (a_\nu,
b_\nu, c_\nu, d_\nu, e_\nu)^T$, we have ${\cal M}_{\nu\nu LL'} V_\nu =
0$, yielding
\beq
a_\nu = \dfrac{1}{\sqrt{1 + \dfrac{m^2_D}{m^2_S} + \dfrac{m^2_2}{2M^2} + \dfrac{{m'_2}^2}{{M'}^2}}} ~,~~
b_\nu = -\dfrac{m_D}{m_S}~a_\nu ~,~~
c_\nu = -\dfrac{m_2}{\s M}~a_\nu ~,~~
d_\nu = -\dfrac{m'_2}{M'}~a_\nu ~,~~
e_\nu = 0 ~.
\label{nucoeffs}
\eeq

For the charged leptons, the mass terms take the following form:
\beq
-\(\bma {\bar \tau}^- & {\bar L}^- & {\bar L}'^-\ema\)_L {\cal M}_{\tau LL'}
\(\bma \tau^- \\ L^- \\ L'^- \ema\)_R ~~,~~~~
{\rm with} ~~~~
{\cal M}_{\tau LL'} =
\(\bma
m_1 & m_2 & -m'_2/\s \\
0   & M   & 0 \\
0   & 0   & M'
\ema\) ~.
\eeq
The masses and mixings relevant for the physical left-handed
(right-handed) states are found by diagonalizing ${\cal M}_{\tau
  LL'}{\cal M}^T_{\tau LL'}$ (${\cal M}^T_{\tau LL'} {\cal M}_{\tau
  LL'}$). We have
\bea
(\tau^-_L)_{phys} & = & a_{L\tau} \tau^-_L + c_{L\tau} L^-_L + d_{L\tau} L^{\prime -}_L ~, \nn\\
(\tau^-_R)_{phys} & = & a_{R\tau} \tau^-_R + c_{R\tau} L^-_R + d_{R\tau} L^{\prime -}_R ~.
\label{tauLmodel(iv)}
\eea
Defining $V_{L\tau} \equiv (a_{L\tau}, c_{L\tau}, d_{L\tau})^T$ and
$V_{R\tau} \equiv (a_{R\tau}, c_{R\tau}, d_{R\tau})^T$, and assuming
that the lightest eigenvalue for the lepton mass matrix is $m_\tau$,
the coefficients are found from
\beq
{\cal M}_{\tau LL'}{\cal M}^T_{\tau LL'} V_{L\tau} = m^2_\tau V_{L\tau} ~~,~~~~
{\cal M}^T_{\tau LL'} {\cal M}_{\tau LL'} V_{R\tau} = m^2_\tau V_{R\tau} ~.
\eeq
This yields
\bea
& a_{L\tau} = \dfrac{1}{\sqrt{1 + \dfrac{m^2_2 M^2}{(M^2 - m^2_\tau)^2} + \dfrac{{m'_2}^2 {M'}^2}{2({M'}^2 - m^2_\tau)^2}}} ~~,~~~~
c_{L\tau} = -\dfrac{m_2 M}{M^2 - m^2_\tau}~a_{L\tau} ~~,~~~~
d_{L\tau} = \dfrac{m'_2 M'}{\s({M'}^2 - m^2_\tau)}~a_{L\tau} ~, & \nn\\
& a_{R\tau} \simeq 1 ~~,~~~~
c_{R\tau} = O(m_\tau v / M^2) ~~,~~~~
d_{R\tau} = O(m_\tau v / M^2)  ~. &
\label{taucoeffs}
\eea
$c_{R\tau}$ and $d_{R\tau}$ are both $O(10^{-4})$. Thus, to a good
approximation, there is no mixing in the right-handed sector.

\end{document}